\documentstyle[prc,aps,floats,epsf]{revtex}

\begin{document}

\draft

\title{Pion-Nucleus Scattering at Medium Energies 
with\\ Densities from Chiral Effective Field Theories}

\author{ B. C. Clark,  R. J. Furnstahl, L.  Kurth Kerr, and John Rusnak}
\address{
Department of Physics, The Ohio State University, Columbus, Ohio 43210-1106}
\author{S. Hama}
\address{Hiroshima University of Economics, Hiroshima 731-01, Japan}

\date{February, 1998}

\maketitle
\begin{abstract}
 Recently developed chiral effective field theory models provide 
 excellent descriptions of the bulk characteristics of finite nuclei, 
 but have not been tested with other observables.  In this work, 
 densities from both relativistic point-coupling models and 
 mean-field meson models are used in the analysis of meson-nucleus 
 scattering at medium energies.  Elastic scattering observables for 790 
 MeV/$c$ $\pi^{\pm}$ on $^{208}$Pb are calculated in a relativistic 
 impulse approximation, using the Kemmer-Duffin-Petiau formalism to 
 calculate the $\pi^{\pm}$ nucleus optical potential.  
\end{abstract}

The concepts and methods of effective field theory (EFT)
\cite{WEINBERG79,GEORGI93,WEINBERG95}
have recently elucidated the 
successful nuclear phenomenology of relativistic field theories
of hadrons, called quantum hadrodynamics (QHD) 
\cite{SEROT86,SEROT92,FURNSTAHL97,SEROT97}.
The EFT framework shows how QHD models can be consistent with
the symmetries of quantum chromodynamics (QCD) 
and can be extended to accurately reproduce its low-energy features.
The EFT perspective 
accounts for the success of relativistic mean-field 
models and provides an
expansion scheme at the mean-field level and
for going beyond it \cite{FURNSTAHL97,RUSNAK97}.

A practical outcome of these EFT studies has been new sets of
relativistic mean-field models, with parameters determined by
global fits to bulk nuclear observables.
Here we make the first independent tests of densities from
these models by using them as inputs to relativistic impulse
approximation (RIA) calculations of elastic $\pi^{\pm}$ nucleus scattering.  
At energies above the $\Delta$ resonance 
we expect the impulse approximation 
to reproduce experiment at forward angles, 
since the elementary amplitudes
incorporate the dominant effects of intermediate $\Delta$'s while
medium modifications due to the $\Delta$ should be
small.

In Ref.~\cite{FURNSTAHL97}, an effective hadronic lagrangian
consistent with the symmetries of QCD and intended for application
to finite density systems was constructed.
The goal was to test a systematic expansion
for low-energy observables,
which included the effects of hadron compositeness and the constraints
of chiral symmetry.
The degrees of freedom are (valence) nucleons, pions, and the
low-lying non-Goldstone bosons.
A scalar-isoscalar field with a mass of roughly 500\,MeV was also included
to simulate the exchange between nucleons of two correlated pions in this
channel.
The lagrangian was expanded in powers of the fields and their
derivatives, with the terms organized using Georgi's ``naive
dimensional analysis'' \cite{GEORGI84b,GEORGI93,FRIAR96}.

The result is
a faithful representation of low-energy, non-strange QCD, as long as all 
nonredundant terms consistent with symmetries are included.
In addition, the mean-field framework provides a means of
approximately including  higher-order many-body and loop effects,
since the scalar and vector meson fields play the role of
auxiliary Kohn--Sham potentials in relativistic density functional 
theory \cite{SEROT97}.
Fits to nuclear properties at the mean-field level
showed that the effective lagrangian could be truncated at the first
few powers of the fields and their derivatives,
with natural [$O(1)$] coefficients for each term.

An analogous study was made of relativistic ``point-coupling'' (PC) models 
\cite{MADLAND} in 
Ref.~\cite{RUSNAK97}.
In these models, non-Goldstone mesons 
(such as the $\sigma$ and $\omega$) are not included explicitly.
Instead one has an expansion of the nucleon scalar and vector potentials
in powers and derivatives of local nucleon scalar and vector densities.
As in Ref.~\cite{FURNSTAHL97}, an effective lagrangian
consistent with chiral symmetry was constructed and fit to bulk nuclear
observables.

These studies generated parameter sets from 
global fits to the binding energies, the charge form 
factors, and spin-orbit splittings of the doubly magic nuclei.
Despite the many observables, the constants in the EFT models are 
underdetermined and many distinct parameter sets with low $\chi^{2}$  
were found \cite{FURNSTAHL97,RUSNAK97,RUSNAK97b}.
The corresponding models have what appear to be significant differences. 
In contrast to conventional models used as input to past impulse approximation
calculations, for which the effective nucleon mass
$M^{*}/M$ at nuclear matter equilibrium was always close to 
0.6,
the new models feature a wide range of effective masses,
some with $M^*/M$ as high as $0.74$.
In addition, there are 
differences in predictions for observables not included in the 
fits, such as neutron radii.

One might hope to distinguish these models and constrain the
parameters by using the predicted
nuclear densities to calculate other observables.
Recent high-quality data for elastic $\pi^{\pm}$ nucleus scattering 
provides such an opportunity.
At the same time, we can test the limits of the relativistic
impulse approximation approach to hadron-nucleus reactions.

To do so, we apply
the Kemmer-Duffin-Petiau (KDP) equation \cite{kemm,duff,pet}, 
which has been used in the analysis
of meson-nucleus scattering at medium energies 
in Refs.~\cite{prl,ku94,george}. 
The KDP equation is similar in form to
the Dirac equation, so it can form the basis of
a relativistic impulse approximation  approach analagous to that 
 used successfully in the treatment
of elastic proton-nucleus scattering \cite{ria1,lanny}.
This similarly is apparent from the KDP free particle equation  

\begin{equation}
\left(\,i\beta^{\mu}\partial_{\mu} - m\,\right)\phi = 0 \ ,
    \label{eq:one}
\end{equation}
where the $\beta^{\mu}$ obey~\cite{kemm} 
\begin{equation}
\beta^{\mu}\beta^{\nu}\beta^{\lambda} + \beta^{\lambda}\beta^{\nu}
\beta^{\mu} = g^{\mu\nu}\beta^{\lambda} + g^{\lambda\nu}\beta^{\mu} \ .
  \label{eq:two}
\end{equation}
The algebra generated by the four $\beta^{\mu}$ has three irreducible 
representations of dimension one, five, and ten, corresponding to trivial, 
spin-0, and spin-1 wave equations, respectively.  In describing spin-0 
pions we make use of the 5-dimensional representation in which the 
$\beta^{\mu}$ are $5 \times 5$ matrices and $\phi$ is a 5-component spinor.  

In order to apply the KDP 
formalism to $\pi^{\pm}$ nucleus scattering, an interaction potential $U$ is
introduced in Eq.~(\ref{eq:one}). 
As discussed in Ref.~\cite{prl}, the most general form for $U$ contains 
two scalar, two vector, and two tensor terms.  The tensor potentials
 are omitted to 
avoid noncausal effects~\cite{gw}.  For the spin-zero case 
the scalar operators are the unit operator $I$, and the 
$5 \times 5$ operator $P$ whose elements are all zero except
the (1,1) element.  $P$ therefore acts as a projection operator onto the 
first component of the KDP spinor $\phi$.  The vector 
operators are $\beta^{\mu}$ and 
$\tilde\beta^{\mu} = P\beta^{\mu} - \beta^{\mu}P$.  
Following the same procedure in constructing the optical model
potentials as in the RIA~\cite{ria1},  the 
$\pi$-nucleon amplitudes are expressed in an invariant form.
  We begin by making 
the following choice for the $\pi$-nucleon $t$ matrix:
\begin{equation}
t = I_N I t'_s + I_N P t_s + \gamma_{\mu}\beta^{\mu}t'_v 
+ \gamma_{\mu}\beta^{\mu}P t_v \ ,
  \label{eq:three}
\end{equation}
with $I_N$ and $\gamma_{\mu}$ being the unit and Dirac matrices for the 
nucleon.  We then equate matrix elements of the empirical $\pi$-nucleon
scattering amplitude
\begin{equation}
F(q) = f(q) + {\bf \sigma}\cdot{\bf n}\; g(q) \ ,
  \label{eq:four}
\end{equation}
taken between Pauli spinors for the nucleon with matrix elements of the 
invariant $t$ matrix taken between Dirac and Kemmer free-particle 
spinors.

\begin{figure}[t]
 \setlength{\epsfxsize}{4.1in}
  \centerline{\epsffile{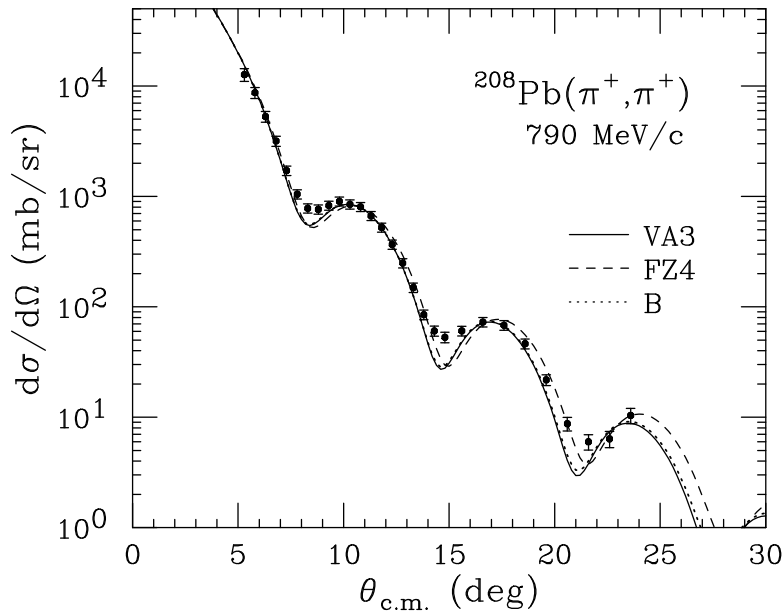}}
\vspace{.1in} 
\caption{\small Differential cross section predictions for 
$^{208}\mbox{Pb}(\pi^+,\pi^+)$ at 790 MeV/c 
calculated using two EFT mean-field densities (VA3 
\protect\cite{RUSNAK97b} and FZ4 \protect\cite{RUSNAK97})
and a relativistic hartree density (B) from Ref.~\protect\cite{fpw}.
The data are from Ref.~\protect\cite{tak-proc,tak2}.}
 \label{fig:piplus}
\end{figure}

In Ref.~\cite{prl}, we limited $t$ to only two of the terms in 
Eq.~(\ref{eq:three}), 
one scalar and one vector.  This gives four possibilities for the 
form of $t$.  For pion scattering the case that
 has the projection operator $P$  in both the scalar 
and vector terms is the only case that gives good agreement with medium 
energy $\pi^{\pm}$ nucleus elastic scattering data.
A matrix $K$ relates the amplitudes  $t$ and $F$ as given below,
\begin{equation}
{t_s \choose t_v} = -\frac{2\pi\sqrt{s}}{Mm}\,K^{-1}\,{f \choose g},
\end{equation}
where $M$ is the nucleon mass, $m$ is the meson mass and $\sqrt{s}$ is
the total meson-nucleon energy. The matrix $K$ is given explicitly in 
Ref.~\cite{prl}.
In contrast to the first-order nonrelativistic impulse approximation, NRIA,
considered in Ref.~\cite{prl},
 where only $f$ contributes to the optical potential,
 both $f$ and $g$ contribute to the first order KDP-RIA potentials.
The scalar and vector optical potentials are  
constructed by folding the invariant amplitudes in Eq.~(\ref{eq:three})
with the nuclear 
densities $\rho$ from the chiral EFT mean-field  models described earlier.
The potentials are written,
\begin{equation}
U_{s,v}(r) = -\frac{P_{lab}}{(2\pi)^2 m}\sum_{i=p,n}\int \,
\frac{dq^3}{(2\pi)^3}\,e^{i{\bf q} \cdot {\bf r}}\,
t_{s,v}^{(i)}(q)\,\tilde \rho_{s,v}^{(i)}(q) \ .
  \label{eq:six}
\end{equation}
In all calculations presented here, 
the summer 1997 amplitudes of Ref.~\cite{arndt} are used.

\begin{figure}[t]
 \setlength{\epsfxsize}{4.1in}
  \centerline{\epsffile{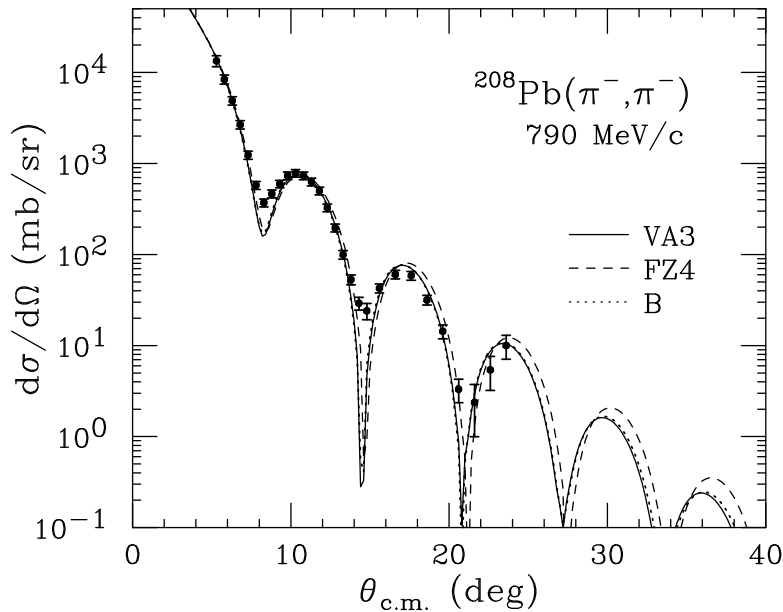}}
\vspace{.1in} 
\caption{\small Differential cross section predictions for 
$^{208}\mbox{Pb}(\pi^-,\pi^-)$ at 790 MeV/c 
calculated using two EFT mean-field densities (VA3 
\protect\cite{RUSNAK97b} and FZ4 \protect\cite{RUSNAK97})
and a relativistic hartree density (B) from Ref.~\protect\cite{fpw}.
The data are from Ref.~\protect\cite{tak-proc,tak2}}
 \label{fig:piminus}
\end{figure}

The KDP-RIA predictions for
790 MeV/c $^{208}Pb(\pi^{\pm},\pi^{\pm})$ elastic scattering differential
cross sections are shown in  
Figs.~\ref{fig:piplus} and \ref{fig:piminus}.
A large number of point-coupling and mean-field meson model densities
were tested, but we show results for only two:  the meson model VA3 
from Ref.~\cite{RUSNAK97b} and the point-coupling model FZ4 from
\cite{RUSNAK97}, which are representative of the range in the models.

These figures show that both densities give good predictions of the 
experimental data, validating the impulse approximation at these
energies.
The differences between the curves
can be traced to the difference in the neutron
distributions;
the neutron rms radii of the two models shown differ by
0.18\,fm, while the proton rms radii differ by only 0.01\,fm. 
Specifically, the 
proton and neutron rms radii are 5.438 fm and 5.770 fm in
the VA3 model 
but 5.428 fm and  5.588 fm in the FZ4 model. 
A sensitivity to the difference in neutron rms radii
has been noted as well in analysis of this data by Takahashi~\cite{tak2}.

Other differences in these densities and others not shown, such as the
range in $M^{*}/M$, are not singled out by this analysis.
A good reproduction of the nuclear charge densities appears to be
sufficient.
Thus, it is not surprising  
that the older mean-field Hartree densities (such as set B \cite{fpw} in the
figures)
give comparable agreement with 
the experimental data.

In summary, we have applied the KDP-RIA formalism to
obtain optical potentials for use in the KDP equation, using mean-field
densities from chiral effective field theory models.  These
simple folding model potentials give quite reasonable agreement with the
experimental 
790 MeV $^{208}Pb(\pi^{\pm},\pi^{\pm})$ elastic scattering cross sections.
A comparison of results from different effective field theory models 
shows some sensitivity to differences in the neutron rms radii 
obtained from the point coupling model and the meson models, but
do not otherwise constrain the many chiral EFT parameter sets.
Analogous studies of proton-nucleus scattering will be discussed
in a forthcoming publication. 

\bigskip

We thank B. D. Serot for useful discussions.
This work was supported in part by the National Science Foundation
under Grants No.\ PHY--9511923 and PHY--9258270,
and The State of Ohio Supercomputing Center.

\end{document}